\documentstyle[11pt,aaspp4]{article}

\def\arcs{\ifmmode {^{\scriptscriptstyle\prime\prime}}
          \else $^{\scriptscriptstyle\prime\prime}$\fi}
\def\arcm{\ifmmode {^{\scriptscriptstyle\prime}}
          \else $^{\scriptscriptstyle\prime}$\fi}
\newdimen\sa  \newdimen\sb
\def\parcs{\sa=.07em \sb=.03em
     \ifmmode $\rlap{.}$^{\scriptscriptstyle\prime\kern -\sb\prime}$\kern -\sa$
     \else \rlap{.}$^{\scriptscriptstyle\prime\kern -\sb\prime}$\kern -\sa\fi}
\def\parcm{\sa=.08em \sb=.03em
     \ifmmode $\rlap{.}\kern\sa$^{\scriptscriptstyle\prime}$\kern-\sb$
     \else \rlap{.}\kern\sa$^{\scriptscriptstyle\prime}$\kern-\sb\fi}
\def\pdeg{\ifmmode $\setbox0=\hbox{$^{\circ}$}\rlap{\hskip.11\wd0 .}$^{\circ}
          \else \setbox0=\hbox{$^{\circ}$}\rlap{\hskip.11\wd0 .}$^{\circ}$\fi}
\def\gtorder{\mathrel{\raise.3ex\hbox{$>$}\mkern-14mu
             \lower0.6ex\hbox{$\sim$}}}
\def\ltorder{\mathrel{\raise.3ex\hbox{$<$}\mkern-14mu
             \lower0.6ex\hbox{$\sim$}}}

\newcommand{\kms}{\mbox{ km~s$^{-1}$}}

\newcommand{\etal}{\mbox{ et~al.}}

\lefthead{}
\righthead{}

\def\ntot{4353}
\def\ngal{4192}
\def\area{2.12}

\begin{document}

\title{The K-Band Galaxy Luminosity Function\footnote{This publication makes use of data products from the Two Micron 
  All Sky Survey (2MASS), which is a joint project of the University of Massachusetts and the Infrared Processing 
  and Analysis Center/California Institute of Technology, funded by the National Aeronautics and Space Administration 
  and the National Science Foundation. }\footnote{This research has made use of the NASA/IPAC Extragalactic 
  Database (NED) which is operated by the Jet Propulsion Laboratory, California Institute of Technology, under 
  contract with the National Aeronautics and Space Administration.}} 

\author{C. S. Kochanek, M. A. Pahre\footnote{Hubble Fellow.}, E. E. Falco, J. P. Huchra, J. Mader,}
\affil{Harvard-Smithsonian Center for Astrophysics \\
       60 Garden Street \\
       Cambridge, MA 02138 \\
       ckochanek, mpahre, efalco, jhuchra, jmader@cfa }
\author{T.H Jarrett, T. Chester, R. Cutri,}
\affil{ Infrared Processing and Analysis Center, MS~100-22, \\
        California Institute of Technology \\
        Pasadena, CA 91125  \\
        jarrett, tchester, roc@ipac.caltech.edu }
\author{and S.E. Schneider}
\affil{ Department of Astronomy \\
        University of Massachusetts\\ 
        Amherst, MA 01003 \\
        schneider@messier.astro.umass.edu }

\begin{abstract}
We measured the K-band luminosity function  using a complete sample of $\ngal$ 
morphologically-typed 2MASS galaxies with $\mu_{K_s}=20$~mag/arcsec$^2$
isophotal magnitudes $7 < K_{20} < 11.25$~mag 
spread over $\area$~str.  Early-type ($T \leq -0.5$) and late-type ($T > -0.5$)
galaxies have similarly shaped luminosity functions, $\alpha_e=-0.92\pm0.10$ and
$\alpha_l=-0.87\pm0.09$.  The early-type galaxies are brighter, $M_{*e}=-23.53\pm0.06$~mag
compared to $M_{*l}=-22.98\pm0.06$~mag, but less numerous, $n_{*e}=(0.45\pm0.06)\times 10^{-2}h^3$~Mpc$^{-3}$
compared to $n_{*l}=(1.01\pm0.13)\times 10^{-2}h^3$~Mpc$^{-3}$ for $H_0=100 h$~km~s$^{-1}$~Mpc$^{-1}$, 
such that the late-type galaxies slightly dominate the K-band luminosity density,
 $j_{late}/j_{early}=1.17\pm0.12$. 
Our morphological classifications are internally consistent, consistent
with previous classifications and lead to luminosity functions unaffected by the
estimated uncertainties in the classifications.  These luminosity functions accurately 
predict the K-band number counts and redshift distributions for $K \ltorder 18$~mag, 
beyond which the results depend on galaxy evolution and merger histories.
\end{abstract}


\section{Introduction}

The luminosity function (LF) of galaxies, its parameters, dependence on galaxy type,
and evolution are fundamental to observational cosmology and the theory of
galaxy formation.  Most 
existing estimates of the luminosity function are based on redshift surveys of 
galaxies selected from blue photographic plates (CfA/CfA2, Davis \& Huchra 1982,
de Lapparent, Geller \& Huchra 1989, Geller \& Huchra 
1989, Marzke et al. 1994ab; SSRS2 da Costa et al. 1994, 1998, Marzke et al. 1998;
APM, Loveday et al. 1992; ESO Slice, Vettolani et al. 1997, Zucca et al. 1997; 
2dFGRS Folkes et al. 1999, Slonim et al. 2000).  
The luminosity function derivations are usually
based on samples of $\sim 5000$ galaxies.  Blue surveys emphasize galaxies 
with active star formation, are sensitive to both Galactic and internal 
extinction, and those based on photographic plates usually have large photometric 
uncertainties (0.2--0.4~mag).
Deep, blue-selected surveys must also include strong, type-dependent 
K-corrections.  The only ongoing blue survey is the 2dFGRS of 250,000
galaxies.

Most recent surveys have shifted to selecting galaxies in the red, which somewhat reduces
the effects of extinction and leads to samples less influenced by recent star
formation.  The Century Survey (Geller et al. 1997) used objects selected from 
red photographic plates with the photometry recalibrated by R$_c$-band drift scans, 
while the Las Campanas Redshift Survey (hereafter the LCRS, Shectman et al. 
1996, Lin et al., 1996, Bromley et al. 1998) selected the galaxies from 
Gunn~r-band drift scans calibrated to approximate Kron-Cousins R$_c$.  The 
Sloan Digital Sky Survey (SDSS) will obtain a surface-brightness limited
sample to $r'=17.7$~mag with approximately $10^6$ galaxies (see York et al. 2000).

Infrared galaxy surveys have smaller systematic uncertainties than optical
galaxy surveys.  They are almost immune to both Galactic and internal 
extinction, and the K-corrections and luminosity per unit stellar mass are 
nearly independent of galaxy type (e.g. Cowie et al. 1994, Gavazzi, Pierini \&
Boselli 1996). The determination of infrared 
luminosity functions has proceeded slowly, however, because of the difficulty 
of obtaining large complete samples.  Mobasher, Sharples \& Ellis (1993) 
and Loveday (2000) obtained infrared photometry of optically-selected galaxies
to estimate the infrared LF. Glazebrook et al. (1994, 1995), Gardner et al. 
(1997) and Szokoly et al. (1998) used relatively deep IR surveys of small 
regions, where the faintness of the targets makes it difficult to obtain 
redshifts of the full sample.  De Propris et al. (1998) and Andreon \&
Pello (2000) have also estimated the infrared luminosity function by
constructing volume limited samples of galaxies in the Coma cluster.
The resulting samples are typically 10
times smaller than published optical samples ($\sim500$ rather than
$\sim 5000$ galaxies).

The 2MASS project (Skrutskie et al. 1997) is obtaining a complete infrared map of the sky, 
with a limiting magnitude for its galaxy catalog of $K_s \simeq 13.5$~mag.  Since
2MASS overlaps the existing optical surveys, it is easy to rapidly generate
large, complete infrared redshift surveys.  In this paper we discuss 
an infrared redshift survey overlapping the CfA2 survey and the 
updated Zwicky catalog (UZC, Falco et al. 1999).  To a 
magnitude limit of $K_{20}\leq 11.25$~mag, $\sim90$\% of the galaxies already have 
redshifts and the remainder were obtained as part of our redshift survey.
For the first time we can derive infrared luminosity 
functions from samples of comparable size to that of the published 
optical luminosity functions.  In \S2 we discuss the sample selection,
in \S3 we derive the luminosity function by galaxy type, and in \S4
we compare the results to other estimates of the luminosity function.
In \S5 we use our luminosity functions to predict the properties of
fainter infrared galaxy samples, and in \S6 we summarize our results.

\section{Sample Selection and Data}

We selected $\ntot$ targets from the 2MASS Second Incremental Release Catalog of Extended
Sources using the default K$_s$-band survey magnitudes, $K_{20}$, which is the magnitude 
inside the circular isophote corresponding to a surface brightness of $\mu_{K_s}=20$~mag/arcsec$^2$
(see Jarrett et al. 2000a).  The $K_{20}$ isophotal magnitude is 10--20\% less than the total 
flux, depending on the galaxy type.
We selected all extended sources with $7 \leq K_{20} \leq 11.25$ mag, $\delta \geq 11^\circ$ (J2000) 
and $|b| \geq 20^\circ$, modulated by the actual sky coverage of the release
(see Figure 1).  Although there is no exact correspondence to optical redshift surveys
because of the wide range of optical to infrared galaxy colors, our magnitude limit 
roughly corresponds to $B \ltorder 15$~mag or $R_c \ltorder 14$~mag.  The effective
optical limits are deeper for red early-type galaxies and shallower for blue late-type
galaxies.  Objects which were not galaxies (artifacts, double stars, planetary
nebula, $\cdots$) were removed from the sample by inspection of the 2MASS data flags
and images, the NED databases and digitized POSS-II
\footnote{The Second Palomar Observatory Sky Survey (POSS-II) was made by the California Institute of 
    Technology with funds from the National Science Foundation, the National Geographic Society, the 
    Sloan Foundation, the Samuel Oschin Foundation, and the Eastman Kodak Corporation. The Oschin Schmidt 
     Telescope is operated by the California Institute of Technology and Palomar Observatory.
     }
\footnote{The Digitized Sky 
    Surveys were produced at the Space Telescope Science Institute under U.S. Government grant NAG W-2166. The
    images of these surveys are based on photographic data obtained using the Oschin Schmidt Telescope on 
    Palomar Mountain and the UK Schmidt Telescope. The plates were processed into the present compressed 
    digital form with the permission of these institutions. } 
images of the targets, 
leaving a sample of $\ngal$ galaxies.  We determined the survey area by integrating over the
survey scans (the scans are $8\farcm5\times 6^\circ$).  Note that the lower
boundary of the scanned region actually lies between $11.5^\circ$ and $12^\circ$ in 
Declination.  The scanned regions inside the angular boundaries cover $\Delta\Omega = \area$~str 
which is 2/3 of the area inside the boundaries.  The uncertainties in the survey area
are less than 5\% but they are difficult to estimate precisely because they
depend on the detailed treatment of galaxies near the edges.  A few percent of the area is also
masked by bright stars.  

Since the survey region overlaps the CfA2 redshift survey (Geller \& Huchra 1989) and
the UZC (Falco et al. 1999), almost all the galaxies in the sample had known redshifts.  
We based our redshift catalog on ZCAT (Huchra et al. 1992, 
http://cfa-www.harvard.edu/$\sim$huchra/zcat),
but checked the redshifts against the UZC reanalysis of the CfA/CfA2 redshift survey
and reconciled or corrected any significant disagreements between the two redshift
catalogs.  The galaxies lacking redshifts were primarily elliptical galaxies whose Zwicky
magnitudes were fainter than the UZC magnitude limit and galaxies outside the 
CfA survey area but inside our Galactic latitude limits.  We obtained the missing
redshifts using the FLWO Tillinghast 1.5m telescope, the FAST spectrograph (Fabricant
et al. 1998), and standard reduction procedures (Kurtz \& Mink 1998).  Figure 1 shows an 
Aitoff projection of the galaxy sample.

\begin{figure}[t]
\epsscale{0.6}
\plotone{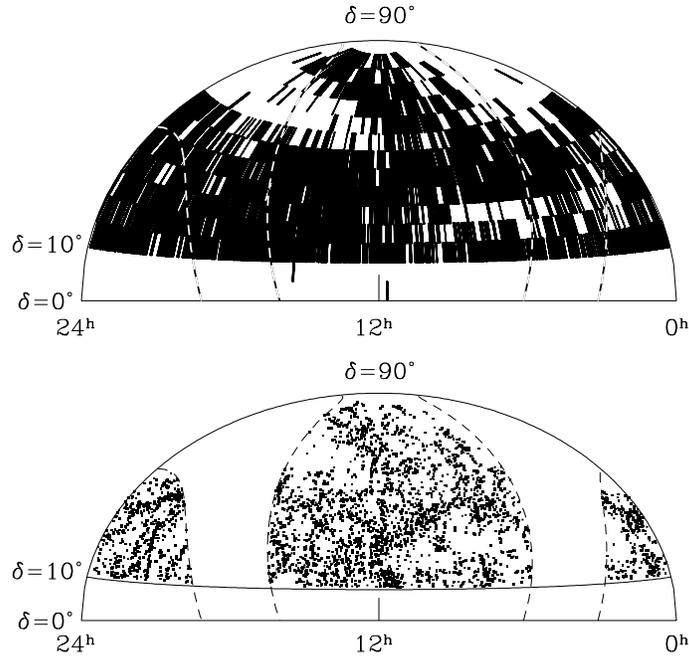}
\caption{
Aitoff projections of the 2MASS scan coverage (top) and the sample galaxies (bottom) in
  equatorial coordinates.  The dashed lines show the $|b| \geq 20^\circ$ Galactic latitude limits
  and the solid line shows the lower declination limit at $\delta \geq  10^\circ$.  
\label{fig-2mass-scans}
  }
\end{figure}

The morphological types of the galaxies are important for studies of galaxy evolution
(e.g. Lilly et al. 1995) and the differences between galaxy environments (e.g. Dressler 1980).  
Our galaxies are relatively nearby,
which allows us to morphologically classify the galaxies.  Of the $\ngal$ galaxies, 
only 1673 have unambiguous types in the RC3 catalog (de Vaucouleurs et al. 1976).  
Each galaxy was visually classified by at least two of the authors (EEF, JPH, CSK
and MAP did the classification) using digitized 
POSS-II images (POSS-I\footnote{The National Geographic Society-Palomar Observatory Sky Atlas (POSS-I) was made 
    by the California Institute of Technology with grants from the National Geographic Society.}
 for the small fraction where POSS-II was unavailable).  
The galaxies were assigned to the classifiers randomly and without information on
the classifications from RC3 or the other classifiers.  We did not, in general,
make use of the full range of fine distinctions in the T-type scale for early-type
galaxies and very late-type galaxies.  Most classifiers used E, E/S0 and S0 for 
early-type galaxies (rather than cE, E, E$+$, S0$-$, S0 and S0$+$), and the
very late-type galaxy classifications (Sd, Sdm, Sm and Im) were not applied 
uniformly.  T-types are more finely grained than we ultimately require, and
our classifications will be internally consistent viewed as the sequence
E, E/S0, S0, S0/a, Sa, Sab, Sb, Sbc, Sc, Scd, Sd$+$later.  Flags were added
for bars (``B''), possible bars (``X''), peculiar morphologies (``pec'') and
evidence for overlapping or interacting neighboring galaxies (``int'').  Our
philosophy for interacting and peculiar galaxies was to assign our best estimate
of the ``intrinsic'' morphology rather than classifying based on the 
transient structures created by the interaction. The
flags were set whenever one classifier assigned it to the galaxy, and they
should be regarded as indicative but not as statistically reliable as the
galaxy types because they were not subject to the same level of inspection.
 
Once the preliminary classifications were complete, we reconsidered the galaxies with
classification ranges covering more than 4 T-types.  These galaxies were reclassified
by all four classifiers with knowledge of all the classifications.  The worst
cases were dominated by interacting galaxies, galaxies with odd star formation
patterns, galaxies classed as ``Irr'' meaning ``peculiar'' rather than ``Im''
($T=10$) in RC3, and the fine grained nature of
the early and very late T-types.   Galaxies with type ranges
greater than 5 T-types were individually discussed.  A limited number of RC3
classifications, largely galaxies classified as ``Irr'', were deleted.  The
final classification was the average T-type of all the classifications.
Figure 2 compares our internal classifications and the RC3 classifications for the 
galaxies in both samples, as a function of the RC3 T-type.  The average difference between
the RC3 T-types and our average T-types is $0.01$ with a dispersion of $1.6$,    
while the average differences between the individual classifiers and RC3 ranged
from $-0.28$ to $+0.24$ with dispersions of $1.8$ T-types.  These statistics
closely resemble the results found by Naim et al. (1995ab) when comparing morphological
classifications of a range of observers and a neural network, and the biases and
scatter are dominated by the very early-type and very late-type galaxies where 
we had not attempted to closely recreate the RC3 classification system.  We
will divide our sample into early-type and late-type galaxies at $T=-0.5$,
so our systematic uncertainties will be dominated by the classification errors
for S0, S0/a and Sa galaxies (see Figure 2).  The galaxy sample is presented 
in Table 1.

\begin{figure}[t]
\epsscale{0.6}
\plotone{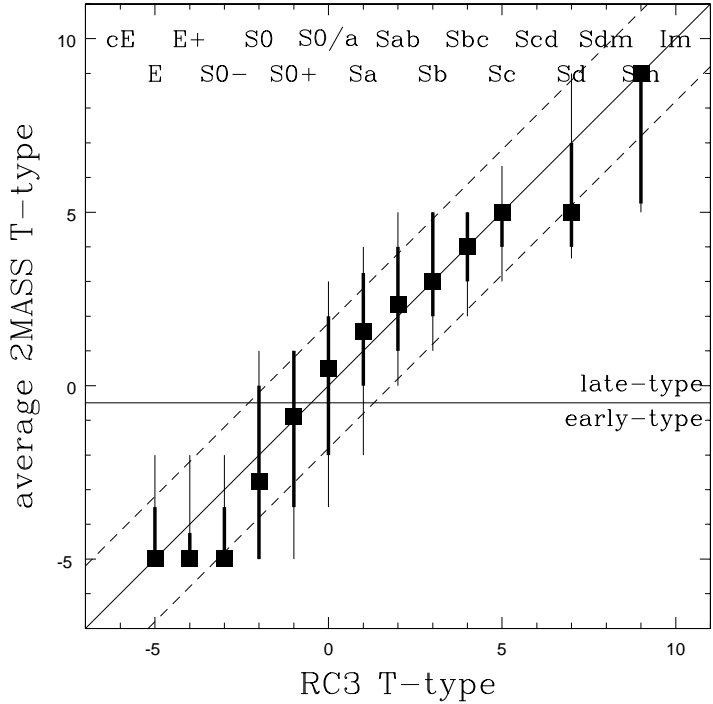}
\caption{
We show the median (point), 1$\sigma$ range (heavy error bars, 68.3\% of galaxies) and 
  2$\sigma$ range (light error bars, 95.4\% of galaxies) of our classifications as a function of
  the RC3 T-type for the 2MASS galaxies found in the RC3 catalog. The dashed lines show the typical
  1.8 dispersion in T-type classifications found by Naim et al. (1995), and the horizontal line
  shows where we break the sample into early-type and late-type galaxies for determining the 
  luminosity function.  Most of the 2MASS classifiers did not use the full range of T-types
  available for early-type galaxies (S0 and earlier) and extremely late-type galaxies 
  (Sd and later), leading to the differences at the edges of the T-type scale.  These differences
  have no affect on our division of the sample into early and late-type galaxies.
\label{fig-class-2mass-rc3}
  }
\end{figure}

The conversion from apparent to absolute magnitude,
\begin{equation}
     M_K = K_{20} - 5 \log (D_L(z)/r_0) - R_K E(B-V) - k(z)
\end{equation} 
has terms for the distance modulus, Galactic extinction $A_K=R_K E(B-V)$
and the K-correction, $k(z)$.  The luminosity distance is $D_L(z)= 6000 h^{-1}(1+z-(1+z)^{1/2})$~Mpc  
for Hubble constant $H_0=100h$~km~s$^{-1}$~Mpc$^{-1}$ and assuming $\Omega_0=1$, although the
particular cosmological model is unimportant given our median redshift of $cz=7000\kms$.  The galaxy 
magnitudes were corrected for Galactic extinction using the extinction maps of Schlegel, Finkbeiner \& Davis (1998) 
and an extinction coefficient of $R_K=0.35$ where $A_K=E(B-V)R_K$ (Cardelli, Clayton \& Mathis 1989).  The
Galactic extinction was less than $E(B-V)=0.03$~mag ($0.14$~mag) for 50\% (95\%) of the sample, and the
maximum extinction was $E(B-V)=0.64$~mag.  Thus, while we include the extinction 
corrections, they are of little importance.  The K$_s$-band K-correction of 
$k(z)= -6.0 \log(1+z)$ is negative, independent of galaxy type, and valid for $z\ltorder0.25$
(based on the Worthey 1994 models).  Unlike most previous estimates of the local luminosity function, 
our intrinsic photometric errors make a negligible contribution to the uncertainties in the LF calculation.
The median error in $K_{20}$ for our sample is 0.03~mag and 90\% of the galaxies have
errors less than $0.04$~mag.  These estimates are verified through repeated scans
of several areas on well-separated nights (Jarrett et al. 2000b).  Our isophotal
magnitudes will have type-dependent differences from integrated magnitudes. 

Given the low redshift of our sample, we need to include corrections for peculiar velocities in
the redshift estimates.  We corrected the heliocentric radial velocities using the local flow model 
of Tonry et al. (2000).  While the Tonry et al. (2000) model is computed using a different cosmology and Hubble 
constant, we use it only as means of estimating the peculiar velocity corresponding to a 
given heliocentric velocity.    For our standard analysis we restrict the sample to galaxies with 
corrected velocities exceeding $cz >2000$~km~s$^{-1}$.  This velocity limit eliminates the Virgo 
cluster from the sample at the price of a significant reduction in the luminosity range of the 
galaxies in the sample.  We
also analyzed the sample down to $cz > 1000$~km~s$^{-1}$, which includes the Virgo cluster and
extends the luminosity function determination to significantly fainter magnitudes at the price
of including galaxies whose luminosities include a significant dependence on the local flow
corrections.

\section{The Luminosity Function}

We used the standard parametric (Sandage, Tammann \& Yahil 1978, hereafter STY)
and non-parametric stepwise maximum-likelihood (SWML, Efstathiou, Ellis \& Peterson 1988) 
methods for determining the shape of the luminosity function, and the Davis \& Huchra (1982) 
minimum variance estimator for determining the absolute number density.  These methods are 
almost universally used for galaxy luminosity function determinations (see Lin et al. 1996
and references therein).  The completeness of the sample and the negligible magnitude errors 
considerably simplify the analysis over most recent studies.
We used the Schechter (1976) parametric model, 
\begin{equation}
   { dn\over dL} = { n_* \over L_*} \left( { L\over L_*} \right)^\alpha \exp(-L/L_*)
\end{equation}
for the STY method fits.  For our standard fits we estimated the luminosity function
using galaxies with flow-corrected velocities $cz > 2000$~km~s$^{-1}$, which excludes
the bulk of the Virgo cluster and restricts us to galaxies with absolute magnitudes
brighter than $M_K < -20.2$~mag.  The density normalization was determined
using the velocity range $2000~\hbox{km~s}^{-1} < cz < 14000~\hbox{km~s}^{-1}$,
the absolute magnitude range $-25\,\hbox{mag} < M_K < -22\,\hbox{mag}$.  We set
the second moment of the correlation function, needed to estimate the effects
of sample variance on the galaxy density, to $J_3=10^4(h^{-1}\hbox{Mpc})^3$ 
(Lin et al. 1996).
We also show fits using galaxies with $cz > 1000$~km~s$^{-1}$, which extends our 
absolute magnitude range to $M_K < -18.7$~mag at the price of increased sensitivity to 
errors in the velocity corrections.  
We have 3878 (4096) galaxies left in the sample with the velocity 
limit $cz > 2000$~km~s$^{-1}$ ($1000$~km~s$^{-1}$).  Our redshift
and magnitude limits also remove almost all the large galaxies ($\gtorder 2'$) 
which have 
unreliable magnitude estimates in the 2MASS Second Incremental Release. The luminosity
function estimation software was tested using synthetic catalogs drawn from 
a Poisson spatial distribution of galaxies selected from Schechter luminosity 
functions.  The SWML binned luminosity functions are presented in Table 2 and
the Schechter function model luminosity functions are presented in Table 3.  

Figure 3 shows luminosity functions for the full sample, the early-type galaxies
and the late-type galaxies using the two different estimation methods. Early-type
galaxies were defined to be all galaxies with $T \leq -0.5$ so that S0/a galaxies
are counted as late-type galaxies and S0$+$ galaxies are counted as early-type
galaxies.  Because the distributions of the T-type classifications are somewhat
quantized, the exact location of the boundary between $-1 < T < 0$ has little effect 
on the results.  The luminosity functions found for the $cz > 2000\kms$ and
$cz >1000\kms$ samples are mutually consistent.  Figure 4 shows the likelihood
contours for the Schechter function $\alpha$ and $M_*$ parameters as compared
to earlier derivations of the infrared luminosity functions.  Note that the 
early-type and late-type luminosity functions have similar shapes, as was also
found in the CfA (Marzke et al. 1994b) and SSRS2 (Marzke et al. 1998) morphologically
classified luminosity functions.  The total luminosity function is steeper than those
of the individual types ($\alpha=-1.09\pm0.06$ rather than $\alpha=-0.87\pm0.09$ or $-0.92\pm01.0$) 
because adding the fainter, more numerous late-type galaxies to the early-type
galaxies makes the summed luminosity function steeper than either of the components.
The values of $\alpha$ and $M_*$ are
strongly correlated, with a dimensionless covariance of 
$C_{\alpha M_*}/(C_{\alpha\alpha} C_{M_* M_*})^{1/2} =0.85$
for all three $cz > 2000\kms$ luminosity functions, as we would expect from the
shapes of the likelihood contours in Figure 4.  The uncertainties in the galaxy
density have similar contributions from sampling errors and changes correlated
with $\alpha$ and $M_*$\footnote{The value of $n_*$ changes with $\alpha$ and
$M_*$ as $n_* = 0.45-0.25\Delta\alpha+0.77\Delta M_*$ for the early-type galaxies
and as $n_*=1.01-0.92\Delta\alpha+2.44\Delta M_* $ for the late-type galaxies
where $\Delta\alpha$ and $\Delta M_*$ are the changes in $\alpha$ and $M_*$
from the maximum likelihood solutions and $n_*$ is in units of $10^{-2}h^3/\hbox{Mpc}^3$.} 
    
We also explored the effects of classification errors on the results.   We first 
examined the effects of simple classification errors using Monte Carlo resampling.  We randomly
selected a new galaxy sample (bootstrap resampling with replacement) including Poisson variations 
in the total number of galaxies.  For each galaxy, we added a 1.8 T-type Gaussian deviate to its
classification before dividing the sample into early-type and late-type galaxy subsamples.  This
random dispersion is a little larger than the 1.6 T-type dispersion between our internal 
classifications and RC3, but matches the dispersion in the morphological classification 
experiments conducted by Naim et al. (1995ab).  The results after repeating the process
100 times are summarized in Table 3, where we present the average parameters and their
dispersions.  These uncertainty estimates will underestimate the uncertainties in the 
absolute density normalization because they include only the Poisson variance in the 
expected number of galaxies without the sample variance due to our survey volume and
larger scale structure.  Aside from the sample variance, the parameter errors and 
correlations estimated by these bootstrap calculations should be more statistically
reliable than those estimated from the likelihood function.  The results are stable to
these statistical errors, since the Schechter function parameters and their bootstrap
uncertainties are consistent with the simpler maximum likelihood estimates.  As we discuss
in Kochanek, Pahre \& Falco (2000), luminosity functions are not stable to even small,
random classification uncertainties when the luminosity function shape depends strongly on the
type (as is found in spectrally-typed luminosity functions like ESP, LCRS and 2dFGRS).
We computed the luminosity density ratio, $j_{late}/j_{early}=1.17\pm0.12$,
using the bootstrap calculations since they include the full variable
covariances and classification uncertainties.  The early-type and late-type
galaxies have nearly equal luminosity densities, although the exact ratio
and the true uncertainties will differ from this estimate
because of the type-dependent corrections between isophotal
and total magnitudes.

Next we explored the sensitivity of the 
results to shifts in the boundary between early-type and late-type galaxies.  In our 
standard LF determination we set the boundary at $T=-0.5$ so that the S0/a galaxies 
(type $T=0$) are counted as late-type galaxies.  In Table 3 we show the 
results of shifting the boundary for early-type galaxies to $T \leq -1.5$
(S0 is the first early-type), $-0.5$ (our standard LF, with S0$+$ is the first early-type), 
$0.5$ (S0/a is the first early-type), and $1.5$ (Sa is the first early-type).  
The parameters $K_*$ and $\alpha$ are insensitive to the boundary shifts, while the
comoving density $n_*$ follows the changes in the relative numbers of galaxies.

Figure 4 and Table 4 compare our Schechter parameter estimates to previous results for the total
infrared luminosity function from Mobasher et al. (1993), Glazebrook et al. (1995), Gardner et al. (1997),
Szokoly et al. (1998), and Loveday (2000).  The sample sizes of these surveys are so much smaller
that their statistical uncertainties dominate any comparison to our results.  All the results are 
mutually consistent given the uncertainties, with the exception of the anomalously high 
density normalization for the Glazebrook et al. (1995) survey.  The uncertainties in the 
Coma luminosity function estimates by de Propris et al. (1988) and Andreon \& Pello (2000)
are significantly larger than for these field surveys.  

\begin{figure}[p]
\epsscale{1.0}
\plotone{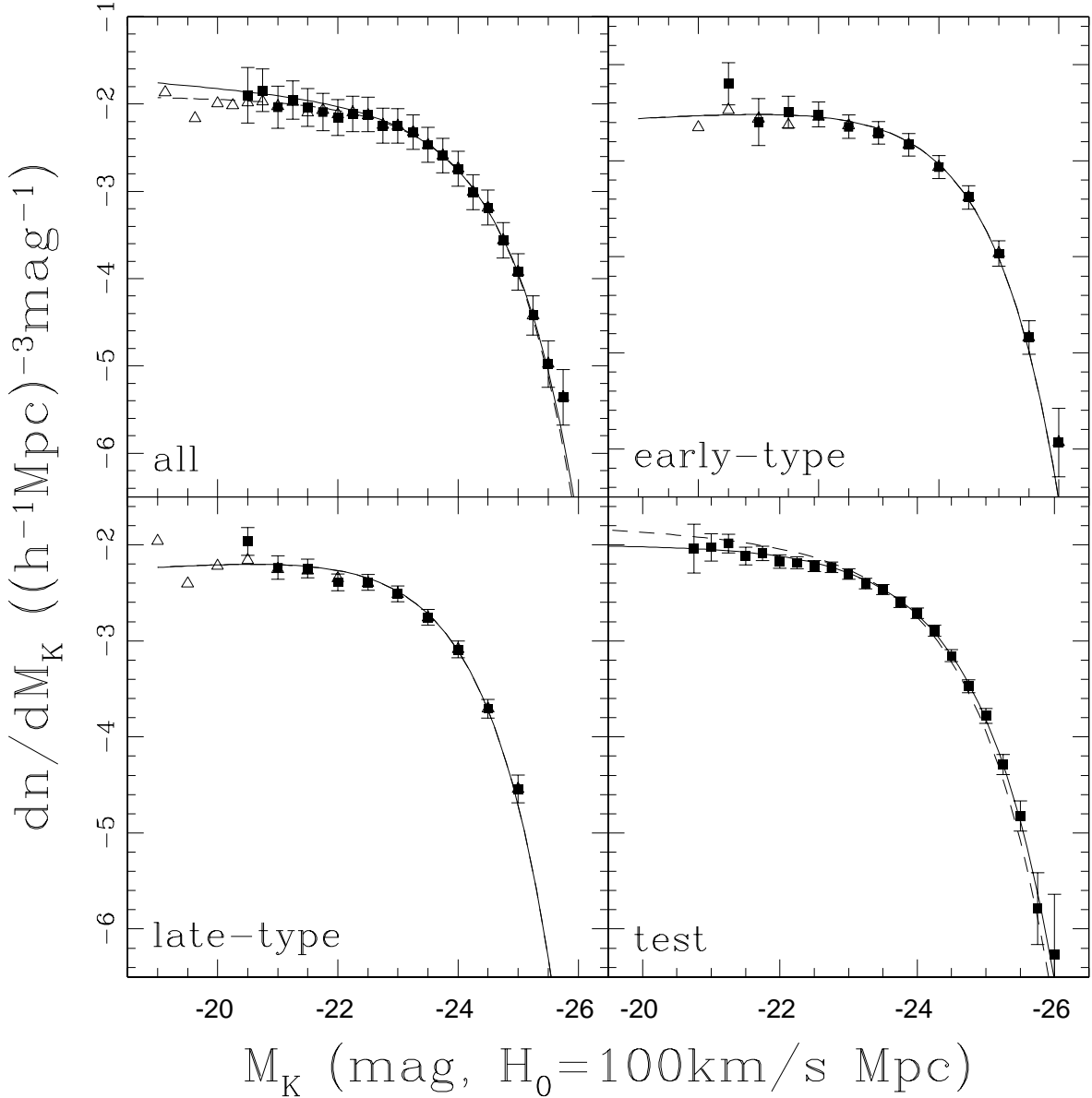}
\caption{
  Luminosity function estimates.  The four panels show the fit to the full sample (top left),
  the early-type galaxy sub-sample (top right), the late-type galaxy sub-sample (bottom left) and
  a Monte Carlo test (bottom right).  The points are the non-parametric SWML model of the 
  luminosity function and the curves are the best fit Schechter functions found with the STY
  method.  The filled squares with error bars and the solid line are for the $cz > 2000\kms$ sample,
  while the open triangles without error bars and the dashed line are for the $cz>1000\kms$ sample. 
  For M$_K\ltorder -21$~mag the symbols for the two samples are superposed.
  The dashed curve in the Monte Carlo test panel is the input luminosity function, which
  was chosen to match the best fit to the full sample.  The error bars are
  highly correlated and include the global uncertainty in the density normalization.  
  Only bins containing at least four galaxies are shown.
\label{fig-lf-estimates}
  }
\end{figure}

\begin{figure}[p]
\epsscale{1.0}
\plotone{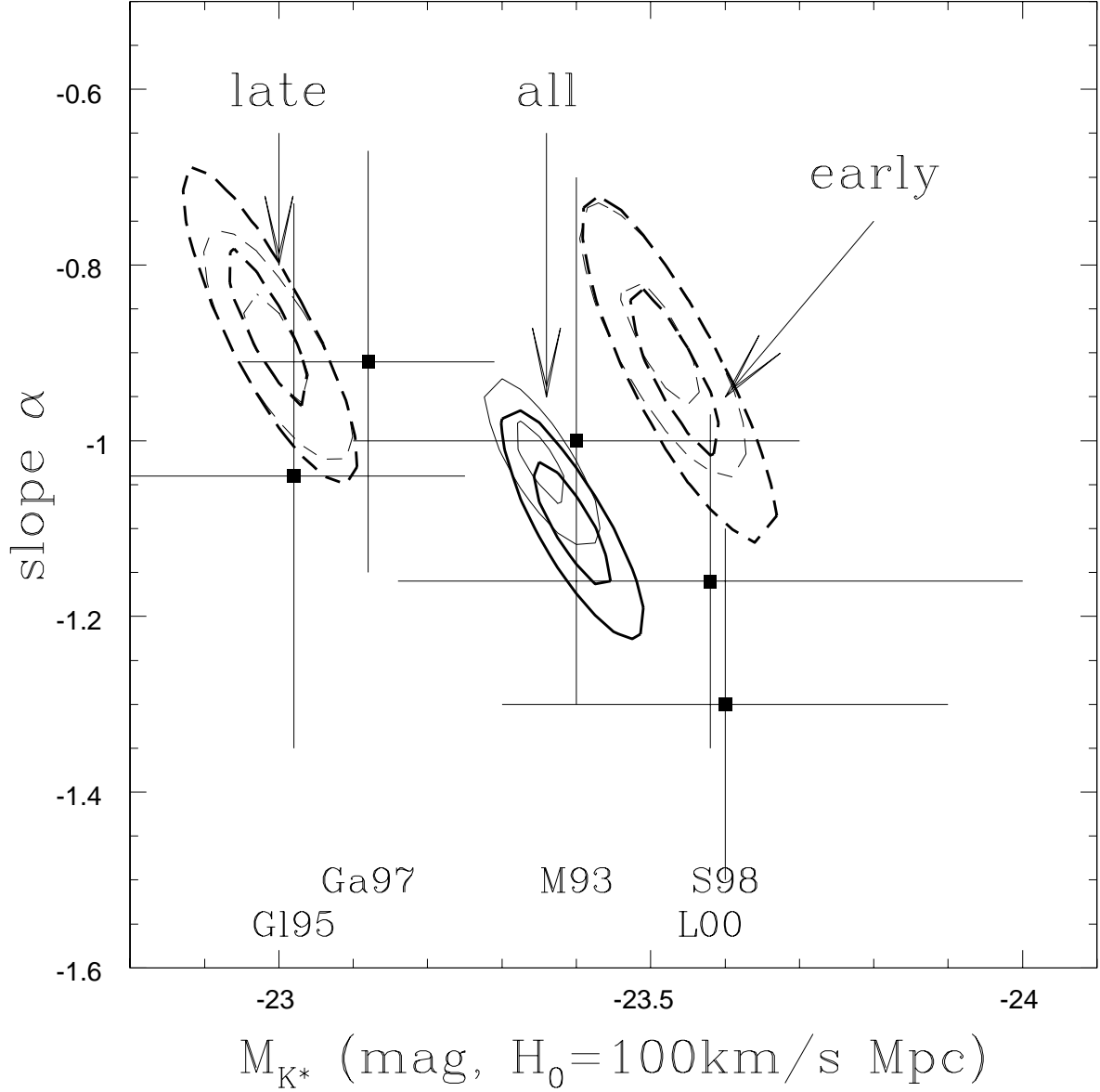}
\caption{
Schechter function parameter likelihoods.  The 1$\sigma$ and 2$\sigma$ likelihood contours 
  for one parameter are shown for the Schechter function parametric fits to the early-type galaxy sub-sample
  (left, dashed), the full sample (middle, solid) and the late-type galaxy sub-sample (right, dashed).
  The heavy contours are for the $cz > 2000\kms$ sample and the light contours are for the $cz>1000 \kms$
  sample.  The points with error bars
  show results from the literature as compiled and standardized by Loveday (2000).  The points are
  (from left to right), Szokoly et al. (1998, S98), Loveday (2000, L00), Mobasher et al. (1993, M93),
  Glazebrook et al. (1995, Gl95), and Gardner et al. (1997, Ga97).  The uncertainties from the 
  present sample are significantly smaller because the sample is complete and
  approximately 10 times larger.
\label{fig-lf-likehoods}
  }
\end{figure}

\begin{figure}[p]
\epsscale{1.0}
\plotone{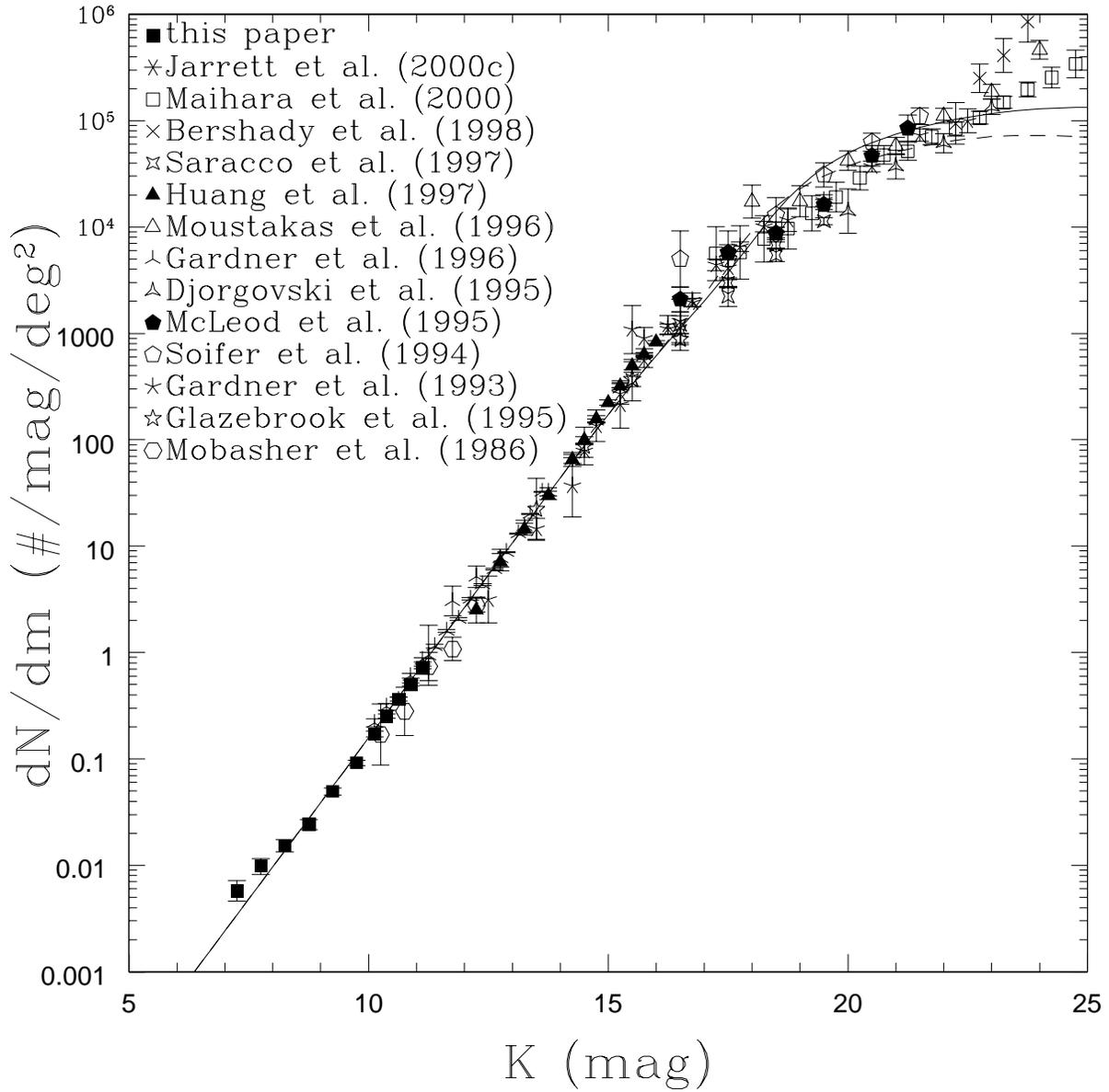}
\caption{
  Differential K-band galaxy number counts.  The points show the results of a wide range of 
  surveys including the number counts of our sample.  The solid (dashed) curve shows the predictions
  for a formation epoch of $z_f=5$ ($z_f=3$).  Our local counts and luminosity
  functions use K$_s$-band isophotal magnitudes (see \S2). We made no corrections for the 
  differences between the K, K$_s$ and K$'$ filters and made no attempt to standardize the
  definitions of the galaxy magnitudes.
  }
\end{figure}

\begin{figure}[p]
\epsscale{1.0}
\plotone{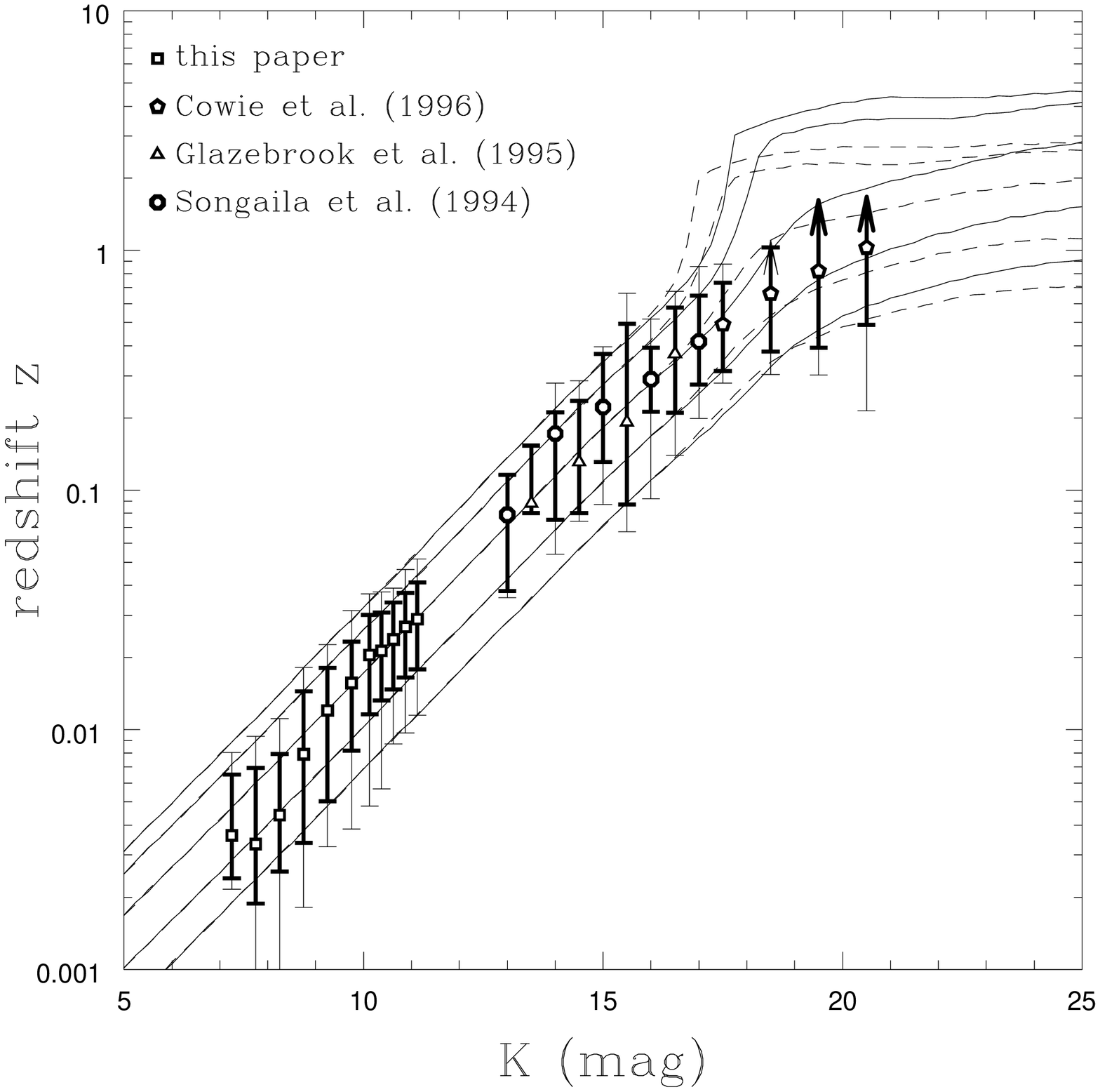}
\caption{
  Redshift distributions predicted by pure luminosity evolution models. The solid (dashed) curves 
  are contours of the redshift distribution for formation epochs of $z_f=5$ ($z_f=3$).
  From top to bottom, 95\%, 84\% (1$\sigma$ above the 
  median), 50\% (the median), 16\% (1$\sigma$ below the median), and 5\% of galaxies are predicted
  to have lower redshifts than the corresponding curve.  The points and error bars show the
  distributions observed in our sample and the fainter samples of Songaila et 
  al.  (1994), Glazebrook et al. (1995)
  and Cowie et al. (1996).  The points correspond to the sample median at each magnitude, the 
  heavy error bars span the 1$\sigma$ region (16\% to 84\% of the sorted sample), and the light 
  error bars span the 5\% to 95\% region.  To construct the sample statistics, unobserved objects 
  were assumed to have the median redshift, and objects with unmeasured redshifts were assumed to 
  lie at high redshift.  An arrow indicates that the upper limit would be due to the objects with 
  unmeasured, but assumed to be high, redshifts, with the tip of the arrow located at the highest 
  measured redshift.  }
\end{figure}

\section{Comparison to Optical Luminosity Functions}

Our infrared luminosity functions are the first large enough to compare directly to the results of recent 
estimates of the luminosity function from optical redshift surveys, which are summarized in Table 5.
The optical luminosity functions, particularly those divided by galaxy type, show 
inconsistencies in their magnitude scales, shapes and density normalizations that are
significantly larger than their formal uncertainties.  In Kochanek et al. (2000)
we show that the luminosity functions defined by spectral types using small aperture fiber
spectrographs (LCRS, ESP, and by extension 2dFGRS and SDSS) have internally inconsistent type 
definitions which 
can severely bias the shapes of the derived luminosity functions.  In essence, the small
spectral apertures sample a varying fraction of the bulge and the disk of spiral galaxies, leading to
flux and luminosity dependent biases between the true and measured spectral types of the
galaxies.  Local, bright, morphologically-typed surveys (this sample, CfA and SSRS2) and large 
aperture spectrally typed surveys (APM survey by spectral type) appear to have self-consistent type 
definitions and similarly shaped luminosity functions for both early and late-type galaxies.
Differences in the surface brightness selection effects of the surveys can also lead to
differences in the shapes of the luminosity functions (e.g. Disney 1976, Huchra 1999, 
Cross et al. 2000).

The luminosity scales ($L_*$ or $M_*$) of the optical surveys 
differ by more than can be explained by
any statistical uncertainties even after including the strong covariances between $\alpha$ and
$M_*$ in Schechter function models of the luminosity function.  For example, the value of 
$M_*$ found for the CfA survey (Marzke et al. 1994ab) is 0.75~mag fainter than the other blue-selected surveys
(APM, ESP, SSRS2 and 2dFGRS).  The current 2MASS catalogs overlap many of the optical surveys, 
which allows us to calculate the extinction and K-corrected average ``color'' $\langle m-K_{20}\rangle$ 
between the optical survey and the 2MASS survey.  This ``color'' includes both the true color
difference of the galaxies and terms due to the different apertures used by the surveys to define
their magnitudes.  Since the intrinsic colors and the 2MASS magnitudes do not depend on the
properties of the optical surveys, we can use the color differences betwen the surveys to test for differences 
in magnitude definitions and type assignments.  We can estimate the average colors for
the APM, CfA, LCRS and SSRS2 surveys, and we present the results in Table 5. 
As we would expect, later-type galaxies are bluer than early-type galaxies, but the color
differences can be significantly smaller than the width of the color distribution.  For 
example, the LCRS emission line and non-emission line samples (Lin et al. 1996) have a
color differences of only $0.15$~mag but a much larger dispersion in the colors of each
type.  The depth of the 2MASS survey is not well matched to that of the LCRS survey, so
our color estimate is dominated by the bright LCRS galaxies (the 2MASS galaxy magnitude
limit of K$_s\simeq 13.5$~mag corresponds to R$_c \simeq 16$~mag while the LCRS survey
is over the range R$_c=15$ to $18$~mag).  The minimal color differences are probably a
symptom of the aperture biases affecting the LCRS spectral classifications, in which bright 
late-type galaxies are misclassified as early-type galaxies (see Kochanek et al. 2000).  
The mismatch of the color differences
between types and the color dispersion for the individual types becomes larger when 
the sample is divided more finely using the spectral clan classification of the LCRS galaxies 
by Bromley et al. (1998).

The colors do not provide a simple explanation for the $M_*$ differences between the
various surveys.  For example, $\langle B-K\rangle \simeq 3.45\pm0.07$~mag for the APM
(B$_J$ mags), CfA (B$_Z$ Zwicky mags) and SSRS2 (B$(0)$ mags) surveys even though the 
characteristic magnitude of the CfA survey is 0.7~mag fainter than the APM and
SSRS2 surveys.  The colors of the early-type galaxies in the CfA and SSRS2 surveys are
very similar, while their colors in the APM survey are $0.3$~mag bluer.  This is consistent
with the incompleteness in the APM morphological classifications being dominated by the more 
distant, red early-type galaxies (see Loveday et al. 1992, Marzke et al. 1994b).  With
the exception of the Sa/Sb galaxies in the CfA survey, which are relatively red and have
an anomalous value for $\alpha$, the late-type galaxies in the three surveys have very
similar colors.  

\section{Comparisons to the Properties of Faint Infrared Samples}

One important use of local luminosity functions is in estimates of the properties of fainter or
higher redshift galaxies.  Here we make some comparisons to the magnitude and
redshift distributions of fainter infrared galaxies using simple evolution models.    
We combined our luminosity functions with Bruzual \& Charlot (1993, GISSEL96 version)
galaxy evolution models assuming an $\Omega_0=0.3$ flat cosmological model and 
$H_0=65$~km~s$^{-1}$~Mpc$^{-1}$ to determine the distances and ages.  We considered
no evolution models (K-corrections only) and evolving models using an ``Sb'' template
for the late-type galaxies (based the star formation history models of 
Guiderdoni \& Rocca-Volmerange 1988) and an 1~Gyr exponential burst 
($SFR\propto \exp(-(t-t_f)/\hbox{Gyr})$) for the early-type galaxies where
the populations formed at $z_f=3$ or $5$.  We predicted the number counts
and redshift distributions of galaxies as a function of magnitude, and compared them
to the available observational data.  All the models are consistent with the number
counts and redshift distributions measured for our low redshift sample (see Figs.
5 and 6), confirming that we have derived luminosity functions consistent with our
data.  Table 6 presents the number counts for our current sample.

It is no surprise that no evolution models with a finite formation epoch are unable
to reproduce either the number counts or the redshift distributions.  The predicted
counts lie below the observations and the predicted redshifts are systematically
lower than observed once $K \gtorder 16$~mag.  Particularly for early-type galaxies,
there is direct evidence that the infrared luminosities evolve significantly by
redshift unity.  Pahre (2000) used the fundamental plane to measure
the amount of surface brightness evolution of cluster early-type galaxies to
$z \simeq 0.5$, de Propris et al. (2000) measured the evolution of the cluster
luminosity functions to $z\simeq 1$, and Kochanek et al. (2000) used the 
fundamental plane of gravitational lenses to measure the surface brightness 
evolution of field early-type galaxies to $z\simeq 1$.  All three estimates
require stellar populations formed in short bursts at $z_f=2$--$5$
rather than no evolution models.
 
Pure luminosity evolution models, where the comoving numbers of galaxies are fixed
but the stellar populations are allowed to evolve, work far better.  Figures 5 
and 6 show that populations formed at $z_f=3$ or $5$ are relatively consistent
with both the number counts and the redshift distributions for $K \ltorder 18$~mag.  
At fainter magnitudes these models begin to have too low a surface density and
too high an average redshift.  The high redshift tail in the distribution is
due to the $z \gtorder 1$  early-type galaxies, which are predicted to be very 
luminous.  Kauffmann \& Charlot (1998) use this disagreement and their
semi-analytic models of galaxy formation to argue that many $L_*$ early-type 
galaxies must be formed from mergers occurring near redshift unity.  Indeed,
crude merger models with $n_* \propto (1+z)^\gamma$, $L_* \propto 1/n_*$
to conserve the total mass and $\gamma\simeq 1$ naturally eliminate the
high redshift tail and increase the number counts of faint galaxies.   

Unfortunately, many of the differences could be created by sample variance rather
than rapid merging.  We can see the 
effects of sample variance in Fig. 6 both at $z\ltorder 0.01$ for our sample and in
the differences between the redshift distributions found by Glazebrook et al. (1995)
and Songaila et al. (1994) at similar apparent magnitudes.  While the differences
in the survey geometries (equal axes versus pencil beam) mean that large scale 
structure affects the survey statistics differently, our comoving volume out to 
$z=0.01$ is 30 times larger than the survey volume of the Cowie et al. (1996) 
fields out to redshift unity.  If we link galaxies in the 
Cowie et al. (1996) fields with velocity differences smaller than $1000\kms$ 
into single ``objects'' (to try to suppress the effects of correlated structures on 
the redshift distribution), the median redshift increases significantly
(by $\Delta z\simeq 0.2$).  Thus, significantly larger redshift samples are
needed to quantitatively test evolutionary models.

\section{Summary}

We have derived the first local infrared galaxy sample whose statistical uncertainties are
comparable to those of local optical galaxy luminosity functions.  We derived both total
and morphologically-typed luminosity functions.  Our morphological types are self-consistent
(see Kochanek et al. 2000) and our luminosity functions are insensitive to random errors
in the classifications and the parameters change as expected when we shift the boundary
between early and late-type galaxies.  Like morphologically-typed optical surveys (CfA
and SSRS2), we find that the luminosity functions of early and late-type galaxies have
similar shapes, $\alpha\simeq -0.9\pm0.1$, in marked contrast to spectrally-typed
optical surveys (ESO Slice, LCRS, 2dFGRS) which usually find that the slope steepens
for late-type galaxies.  Note, however, that in Kochanek et al. (2000) we find that
the spectral classification methods are not self-consistent because of the aperture 
bias created by using a spectroscopic aperture that is much smaller than the galaxies 
being observed.  We used galaxies found in both the optical redshift surveys and the
2MASS survey to estimate the magnitude differences between 2MASS and the optical 
surveys and also between the different optical surveys.  In all surveys, later
type galaxies have bluer optical to infrared colors, but the magnitude differences
cannot fully explain the discrepancies between the magnitude scales of the
luminosity functions.  Our luminosity functions successfully predict the properties of fainter 
infrared samples until $K \gtorder 18$~mag where the models have a significant dependence 
on galaxy evolution and merging histories and the comparison data is probably affected 
by sample variance.  

These results are preliminary, and the sample is still growing rapidly.  In particular,
the survey area complete to the current magnitude limit continues to expand rapidly, and
it is easy to build complete, deeper samples in restricted areas to extend to fainter
absolute magnitude limits.  With complete sky coverage we can use the 2MASS catalog to
probe the relative completeness of redshift surveys and to improve our comparisons 
between the survey magnitude scales.  By combining this with the surface photometry available 
for 2MASS galaxies we can quantitatively explore the effects of surface brightness selection 
effects (e.g. Disney 1976, Sprayberry et al. 1997, Dalcanton et al. 1997, Huchra 1999, Cross et al. 2000) on large
redshift surveys.  As the coverage gaps are eliminated we can look at density 
dependences to galaxy properties.  In Pahre et al. (2000) we derive an improved galaxy 
velocity function ($dn/d\log v$ instead of $dn/dM$) based on the Tully-Fisher and
Faber-Jackson relations derived from the same 2MASS photometry used to derive the
luminosity function.  The velocity function is useful because it determines the
optical depth of the universe to gravitational lensing (see Falco, Kochanek \& Munoz 1998) and 
can be used to probe the evolution of galaxies (see Gonzalez et al. 2000). 

\acknowledgments

\noindent Acknowledgements:  
This research was supported by the Smithsonian Institution.
The authors thank P. Berlind and M. Calkins, the FLWO remote observers,
for obtaining the redshift data, and S. Tokarz for the data reduction.  We thank J. Loveday for
his compilation of infrared luminosity function estimates. 
M.A.P. was supported by Hubble Fellowship grant HF-01099.01-97A from STScI 
(which is operated by AURA under NASA contract NAS5-26555).
 
{}

\def\mc#1{\multicolumn{1}{c}{#1}}
\def\m{\phantom{-}}
\def\s{\phantom{0}}
\def\sp{\phantom{\pm0000}}

\begin{deluxetable}{lrcrrccc}
\tablewidth{0pt}
\tablecolumns{8}
\tablecaption{The Galaxy Sample}
\scriptsize
\tablehead{
\mc{Target}   &\mc{$cz$}   &ref. &\mc{$K_{20}$}   &\mc{T-type} &\mc{Bar} &\mc{Pec?} &\mc{Int?} \\
              &\mc{$\kms$} &code &\mc{mag}        &            &         &          &           }
\startdata
2MJ000009.1+324418 & 10372 &2779 & 10.59 &$ -4.2\pm  1.3$  &   &  &  \\
2MJ000028.8+324656 &  9803 &2700 & 10.89 &$ -4.9\pm  0.1$  &   &  &Y \\
2MJ000038.0+282305 &  8705 &2212 & 10.52 &$  1.5\pm  0.9$  &   &  &  \\
2MJ000044.0+282405 &  8157 &2779 & 11.22 &$  2.3\pm  1.0$  &B &  &Y \\
2MJ000047.0+282407 &  8764 &2779 & 10.33 &$ -4.0\pm  1.5$  &   &  &Y \\
2MJ000058.9+285442 &  6899 &2212 & 11.09 &$  2.0\pm  1.0$  &X &  &  \\
2MJ000103.6+343911 & 12684 &--160& 11.16 &$  2.0\pm  1.0$  & & & \\
2MJ000114.1+344032 & 12953 &2779 & 11.08 &$  3.0\pm  1.0$  &BX &  &Y \\
2MJ000119.7+343132 &  5032 &5502 & 10.62 &$  2.5\pm  1.0$  &   &  &  \\
2MJ000126.7+312600 &  4948 &2700 & 10.26 &$ -3.5\pm  1.5$  &X &Y &Y \\
2MJ000130.0+312630 &  4767 &2212 & 10.42 &$  3.7\pm  1.0$  &   &Y &Y \\
2MJ000138.3+232902 &  4371 &0620 &  9.27 &$  5.1\pm  0.1$  &   &  &Y \\
2MJ000141.9+232944 &  4336 &0620 &  9.94 &$  4.6\pm  0.9$  &   &Y &Y \\
2MJ000246.0+185311 &  7882 &0650 & 10.86 &$  1.7\pm  1.0$  &X &  &  \\
2MJ000309.6+215736 &  6600 &2212 & 10.49 &$  2.5\pm  0.9$  &   &Y &  \\
2MJ000329.2+272106 &  7690 &2700 & 11.06 &$  0.1\pm  1.5$  &X &  &  \\
2MJ000335.0+231202 &  7254 &0668 & 10.97 &$  4.4\pm  1.0$  &X &  &  \\
2MJ000358.7+204502 &  2310 &0658 &  8.57 &$  4.6\pm  0.9$  &   &  &  \\
2MJ000433.7+281805 &  8785 &2700 & 10.61 &$ -3.0\pm  1.5$  &   &  &  \\
2MJ000548.3+272657 &  7531 &0624 & 10.92 &$  3.0\pm  1.0$  &B &  &  \\
2MJ000640.1+260916 &  7552 &2700 & 11.12 &$  2.0\pm  1.6$  &   &  &  \\
2MJ000842.4+372652 &  4389 &0649 & 10.05 &$ -0.3\pm  1.4$  &   &  &  \\
2MJ000932.7+331831 &  4901 &2700 &  9.49 &$ -3.8\pm  1.5$  &   &  &  \\
2MJ001040.8+325858 &  4788 &0668 & 10.64 &$  3.8\pm  1.0$  &B &  &  \\
2MJ001046.8+332110 &  4765 &0668 & 10.49 &$  3.6\pm  1.0$  &X &  &  \\
2MJ001101.0+300307 &  6791 &2212 & 10.18 &$  2.3\pm  1.0$  &X &  &  \\
2MJ001143.1+205832 & 13900 &0300 & 10.99 &$ -3.6\pm  1.8$  &   &  &  \\
2MJ001151.3+330623 & 14058 &2700 & 11.19 &$ -3.6\pm  1.8$  &   &  &  \\
2MJ001215.7+221918 &  7629 &0624 & 10.69 &$  1.0\pm  0.9$  &   &  &  \\
2MJ001218.8+310339 &  4857 &2212 & 10.87 &$  4.8\pm  0.7$  &   &  &  \\
\enddata
\tablecomments{
  The first 30 entries of the catalog.  The redshift $cz$ is the measured heliocentric
  velocity and $K_{20}$ is the isophotal apparent magnitude (see Jarrett et al. 2000a). 
  The ZCAT format reference code for the source of the redshift measurement is given by
  the ``ref. code'' entry (see http://cfa-www.harvard.edu/$\sim$huchra/zcat/zsource.tex).
  The error bar on the T-type classification is the
  standard error based on the scatter in the two or more classifications for the object. 
  In the Bar column we flag objects which at least one classifier flagged as having a 
  full (B) or incipient bar (X).  In the Pec? and Int? columns we flag objects which were
  considered to be peculiar or interacting by at least one classifier.
   }
\end{deluxetable}

\begin{deluxetable}{crccrccrcc}
\tablewidth{0pt}
\tablecolumns{10}
\tablecaption{2MASS Non-Parametric Luminosity Functions}
\scriptsize
\tablehead{
\mc{$M_K$}  &\multicolumn{3}{c}{all} &\multicolumn{3}{c}{early-type} &\multicolumn{3}{c}{late-type} \\
\mc{(mag)}  &$N$   &$\log(n)$   &$\sigma$  &$N$   &$\log(n)$   &$\sigma$         &$N$   &$\log(n)$   &$\sigma$        }
\startdata
$-26.00$ &$  1$ &$ -6.34$ &$  0.66$  	  &$  4$ &$ -5.93$ &$  0.36$  	   \\     
$-25.75$ &$  9$ &$ -5.36$ &$  0.32$  	 \\	          
$-25.50$ &$ 16$ &$ -4.98$ &$  0.27$  	  &$ 37$ &$ -4.84$ &$  0.17$  	  &$  3$ &$ -5.81$ &$  0.45$\\
$-25.25$ &$ 41$ &$ -4.42$ &$  0.23$  	 \\	   
$-25.00$ &$ 94$ &$ -3.92$ &$  0.21$  	  &$160$ &$ -3.97$ &$  0.13$  	  &$ 33$ &$ -4.54$ &$  0.15$\\
$-24.75$ &$169$ &$ -3.56$ &$  0.20$  	 \\	   
$-24.50$ &$308$ &$ -3.19$ &$  0.20$  	  &$389$ &$ -3.38$ &$  0.12$  	  &$173$ &$ -3.71$ &$  0.10$\\
$-24.25$ &$356$ &$ -3.01$ &$  0.20$  	 \\	   
$-24.00$ &$494$ &$ -2.74$ &$  0.20$  	  &$457$ &$ -3.06$ &$  0.12$  	  &$471$ &$ -3.09$ &$  0.09$\\
$-23.75$ &$494$ &$ -2.59$ &$  0.20$  	 \\	   
$-23.50$ &$437$ &$ -2.47$ &$  0.20$  	  &$359$ &$ -2.83$ &$  0.12$  	  &$529$ &$ -2.76$ &$  0.08$\\
$-23.25$ &$401$ &$ -2.32$ &$  0.20$  	 \\	   
$-23.00$ &$327$ &$ -2.25$ &$  0.20$  	  &$210$ &$ -2.71$ &$  0.12$  	  &$428$ &$ -2.51$ &$  0.08$\\
$-22.75$ &$206$ &$ -2.25$ &$  0.20$  	 \\	   
$-22.50$ &$191$ &$ -2.12$ &$  0.20$  	  &$ 94$ &$ -2.65$ &$  0.12$  	  &$261$ &$ -2.39$ &$  0.08$\\
$-22.25$ &$127$ &$ -2.11$ &$  0.20$  	 \\	   
$-22.00$ &$ 65$ &$ -2.15$ &$  0.21$  	  &$ 43$ &$ -2.52$ &$  0.13$  	  &$106$ &$ -2.39$ &$  0.09$\\
$-21.75$ &$ 43$ &$ -2.09$ &$  0.21$  	 \\	   
$-21.50$ &$ 33$ &$ -2.04$ &$  0.22$  	  &$ 16$ &$ -2.49$ &$  0.16$  	  &$ 56$ &$ -2.25$ &$  0.10$\\
$-21.25$ &$ 28$ &$ -1.95$ &$  0.22$  	 \\	   
$-21.00$ &$ 15$ &$ -2.04$ &$  0.24$  	  &$  6$ &$ -2.60$ &$  0.24$  	  &$ 26$ &$ -2.24$ &$  0.12$\\
$-20.75$ &$ 14$ &$ -1.84$ &$  0.24$  	 \\	   
$-20.50$ &$  5$ &$ -1.90$ &$  0.32$  	  &$  5$ &$ -2.20$ &$  0.22$  	  &$ 11$ &$ -1.96$ &$  0.15$\\
$-20.25$ &$  3$ &$ -1.00$ &$  0.21$  	 \\	   
\enddata
\tablecomments{
   The SWML binned luminosity functions as a function of absolute magnitude $M_K$ where
   $\log(n)$ is the logarithm of the comoving density (number/$h^{-3}$Mpc$^3$~mag) and  
   $\sigma$ is its uncertainty.  The late-type and early-type luminosity functions were
   derived using $\Delta M=0.5$~mag bins widths, twice that for the full sample.  The
   errors for the individual bins are very highly correlated and cannot be used 
   directly if the uncertainty weightings are quantitatively important.
   }
\end{deluxetable}

\begin{deluxetable}{llrlll}
\tablewidth{0pt}
\tablecolumns{6}
\tablecaption{2MASS Parametric Luminosity Functions}
\scriptsize
\tablehead{
\mc{Name}   &\mc{Type}   &\mc{$N$}   &\mc{$K_*$} &\mc{$\alpha$} &\mc{$n_*$}           \\
            &            &           &\mc{mag}   &              &$10^{-2}h^3$~Mpc$^{-3}$}
\startdata
Standard           &all   &3878     &$-23.39 \pm0.05$   &$-1.09 \pm0.06$    &$1.16\pm0.10$     \\
                   &late  &2097     &$-22.98 \pm0.06$   &$-0.87 \pm0.09$    &$1.01\pm0.13$     \\
                   &early &1781     &$-23.53 \pm0.06$   &$-0.92 \pm0.10$    &$0.45\pm0.06$     \\
$cz>1000\kms$      &all   &4096     &$-23.35 \pm0.04$   &$-1.02 \pm0.05$    &$1.19\pm0.10$     \\
                   &late  &2244     &$-23.00 \pm0.05$   &$-0.89 \pm0.07$    &$1.00\pm0.12$     \\
                   &early &1852     &$-23.51 \pm0.06$   &$-0.89 \pm0.08$    &$0.46\pm0.06$     \\
Bootstrap          &late  &\mc{$-$} &$-23.02 \pm0.06$   &$-0.96 \pm0.09$    &$0.91\pm0.10$     \\
                   &early &\mc{$-$} &$-23.52 \pm0.05$   &$-0.90 \pm0.09$    &$0.48\pm0.04$     \\
Boundary T$=-1.5$  &late  &2311     &$-22.98 \pm0.06$   &$-0.87 \pm0.09$    &$1.14\pm0.14$     \\
                   &early &1567     &$-23.55 \pm0.07$   &$-0.85 \pm0.11$    &$0.38\pm0.05$     \\
Boundary T$= 0.5$  &late  &1827     &$-22.98 \pm0.06$   &$-0.87 \pm0.10$    &$0.88\pm0.13$     \\
                   &early &2051     &$-23.52 \pm0.06$   &$-0.99 \pm0.09$    &$0.53\pm0.06$     \\
Boundary T$= 1.5$  &late  &1472     &$-23.02 \pm0.07$   &$-0.94 \pm0.10$    &$0.68\pm0.10$     \\
                   &early &2406     &$-23.47 \pm0.06$   &$-0.99 \pm0.08$    &$0.66\pm0.07$     \\
\enddata
\tablecomments{
  The standard model uses a velocity limit $cz > 2000\kms$ and the boundary between early-type and
  late-type galaxies is T$=-0.5$.  The Name column shows the change made to the standard model
  to derive that case's parameters. The Bootstrap case randomly resamples the galaxies with
  replacement, including Poisson variations in the number of galaxies and the addition of 
  random errors to the morphological types (see text).  Its density uncertainties do not
  include the contribution from sample variance due to large scale structure.   We 
  present the Schechter function parameters $K_*$, $\alpha$ and $n_*$ (eqn. 2), and we used
  $H_0=100 h$~km~s$^{-1}$~Mpc$^{-1}$ in estimating $K_*$ and $n_*$.
   }
\end{deluxetable}

\begin{deluxetable}{llllll}
\label{tab-obslog}
\tablewidth{0pt}
\tablecolumns{9}
\tablecaption{Previous Infrared Luminosity Functions}
\scriptsize
\tablehead{
 \mc{Sample}              &$N$   &\mc{$K_*$} &\mc{$\alpha$} &\mc{$n_*$}          &\mc{Type} \\
                          &      &\mc{mag}  &              &$10^{-2}h^3$~Mpc$^{-3}$}
\startdata
Mobasher et al. 1993      & 181  &$-23.4\s\pm0.3\s$  &$-1.0\s\pm0.3\s$   &$1.12 \pm0.16$   &optically selected\\
Glazebrook et al. 1995    & 335  &$-23.02 \pm0.23$   &$-1.04 \pm0.31$    &$2.90 \pm0.70$   &redshift, 37\% complete \\
Gardner et al. 1997       & 567  &$-23.12 \pm0.17$   &$-0.91 \pm0.24$    &$1.66 $          &redshift, 90\% complete\\
Szokoly et al. 1998       & 867  &$-23.6\s\pm0.3\s$  &$-1.3\s\pm0.2\s$   &$1.2\s\pm0.4\s$  &redshift, 31\% complete\\
Loveday 2000              & 345  &$-23.58 \pm0.42$   &$-1.16 \pm0.19$    &$1.2\s\pm0.8\s$  &optically selected\\
de Propris et al. 2000    &      &$-23.3\s\pm0.7\s$  &$-0.8\s\pm0.4\s$   &$--$             &Coma cluster \\
\enddata
\tablecomments{
   Table derived from Loveday (2000). 
   The {\it optically selected} surveys used K-band imaging of galaxies from a complete but optically selected 
   redshift survey, and the {\it redshift} surveys obtained redshifts for objects selected from an infrared
   imaging survey.  De Propris et al. (2000) constructed a volume limited sample in the Coma cluster.
   Mobasher et al. (1993) magnitudes have been adjusted by 0.22~mag 
   due to K-correction differences (see Glazebrook et al. 1995; Gardner et al. 1997).  An aperture
   correction of $-0.30$ is added to the Glazebrook et al. (1995) magnitudes (see Gardner et al. 1997).
   We show the Glazebrook et al. (1995) results for $z<0.2$, which includes only 55 galaxies with redshifts.
   Gardner et al. (1997) contains no estimate for the uncertainties in $n_*$.  The Poisson uncertainties
   are $0.07\times10^{-2}h^3$~Mpc$^{-3}$, but the true error will be dominated by sample variance due to
   the finite survey volume.  All the results are scaled to $H_0=100h$~km~s$^{-1}$~Mpc$^{-1}$.
   }
\end{deluxetable}

\begin{deluxetable}{ccccllllc}
\label{tab-obslog}
\tablewidth{0pt}
\tablecolumns{9}
\tablecaption{Optical Luminosity Functions}
\scriptsize
\tablehead{
 \mc{Survey} &\mc{type} &\mc{$N$}  &\mc{band} &\mc{$M_*$} &\mc{$\langle m-K_{20}\rangle$} &\mc{$\alpha$} &\mc{$n_*$} &Ref\\
             &          &          &          &\mc{mag}   &\mc{mag}                       &         &$10^{-2}h^3$~Mpc$^{-3}$}
\startdata
APM   &all     &1658  &$B_J$ &$-19.50\pm0.13$ &$3.39\pm0.62$  &$-0.97\pm0.15$ &$1.40\pm0.17$   &1\\ 
      &early   & 311  &      &$-19.71\pm0.25$ &$3.73\pm0.47$  &$\m0.20\pm0.35$&                &1\\
      &late    & 999  &      &$-19.40\pm0.16$ &$3.25\pm0.68$  &$-0.80\pm0.20$ &                &1\\
Century   &all &1762  &$R_c$ &$-20.73\pm0.18$ &               &$-1.17\pm0.19$ &$2.50\pm0.60$   &2\\ 
CfA   &all     &9063  &$B_Z$ &$-18.80\pm0.30$ &$3.46\pm0.89$  &$-1.00\pm0.20$ &$4.00\pm1.00$   &3\\ 
      &E       &      &      &$-19.23\pm0.2\s$&$4.10\pm0.65$  &$-0.85\pm0.20$ &$0.15\pm0.04$   &4\\
      &S0      &      &      &$-18.74\pm0.1\s$&$3.95\pm0.65$  &$-0.94\pm0.15$ &$0.76\pm0.20$   &4\\
      &Sa/b    &      &      &$-18.72\pm0.1\s$&$3.79\pm0.56$  &$-0.58\pm0.15$ &$0.87\pm0.22$   &4\\
      &Sc/d    &      &      &$-18.81\pm0.2\s$&$3.34\pm0.64$  &$-0.96\pm0.15$ &$0.44\pm0.11$   &4\\
      &Sm/Im   &      &      &$-18.79\pm0.5\s$&$2.40\pm0.73$  &$-1.87\pm0.20$ &$0.06\pm0.02$   &4\\
ESP   &all     &3342  &$B_J$ &$-19.61\pm0.08$ &               &$-1.22\pm0.07$ &$2.00\pm0.4$    &5\\ 
      &em      &1575  &      &$-19.47\pm0.10$ &               &$-1.40\pm0.10$ &$1.00\pm0.2$    &5\\
      &not-em  &1767  &      &$-19.62\pm0.10$ &               &$-0.98\pm0.09$ &$1.10\pm0.2$    &5\\
LCRS  &all     &18678 &$R_c$ &$-20.29\pm0.02$ &$2.43\pm0.28$  &$-0.70\pm0.03$ &$1.90\pm0.1$    &6\\ 
      &not-em  &11366 &      &$-20.22\pm0.02$ &$2.48\pm0.21$  &$-0.27\pm0.04$ &$1.10\pm0.1$    &6\\
      &em      & 7312 &      &$-20.03\pm0.03$ &$2.32\pm0.35$  &$-0.90\pm0.04$ &$1.30\pm0.1$    &6\\
      &1       & 655  &      &$-20.28\pm0.07$ &$2.54\pm0.17$  &$\m0.54\pm0.14$&$0.034\pm0.003$ &7\\
      &2       &7614  &      &$-20.23\pm0.03$ &$2.50\pm0.18$  &$-0.12\pm0.05$ &$0.71\pm0.06$   &7\\
      &3       &4667  &      &$-19.90\pm0.04$ &$2.44\pm0.26$  &$-0.32\pm0.07$ &$0.99\pm0.13$   &7\\
      &4       &3210  &      &$-19.85\pm0.05$ &$2.33\pm0.32$  &$-0.64\pm0.08$ &$1.15\pm0.21$   &7\\
      &5       &1443  &      &$-20.03\pm0.09$ &$2.18\pm0.34$  &$-1.33\pm0.09$ &$0.84\pm0.22$   &7\\
      &6       & 689  &      &$-20.01\pm0.14$ &$1.89\pm0.38$  &$-1.84\pm0.11$ &$1.31\pm0.78$   &7\\
SSRS2 &all     &5036  &$B(0)$&$-19.43\pm0.06$ &$3.55\pm0.83$  &$-1.12\pm0.05$ &$1.28\pm0.20$   &8\\ 
      &E/S0    &1587  &      &$-19.37\pm0.11$ &$4.07\pm0.58$  &$-1.00\pm0.09$ &$0.44\pm0.08$   &8\\
      &Spiral  &3227  &      &$-19.43\pm0.08$ &$3.32\pm0.81$  &$-1.11\pm0.07$ &$0.80\pm0.14$   &8\\
      &Irr/pec & 204  &      &$-19.78\pm0.45$ &$3.22\pm1.04$  &$-1.81\pm0.24$ &$0.20\pm0.08$   &8\\
2dFGRS&all     &5869  &$B_J$ &$-19.73\pm0.06$ &               &$-1.28\pm0.05$ &$1.69\pm0.17$   &9\\
      &1       &1850  &      &$-19.61\pm0.09$ &               &$-0.74\pm0.11$ &$0.90\pm0.09$   &9\\
      &2       & 928  &      &$-19.68\pm0.14$ &               &$-0.86\pm0.15$ &$0.39\pm0.06$   &9\\
      &3       &1200  &      &$-19.38\pm0.12$ &               &$-0.99\pm0.13$ &$0.53\pm0.08$   &9\\
      &4       &1193  &      &$-19.00\pm0.12$ &               &$-1.21\pm0.12$ &$0.65\pm0.13$   &9\\
      &5       & 668  &      &$-19.02\pm0.22$ &               &$-1.73\pm0.16$ &$0.21\pm0.11$   &9\\
\enddata
\tablecomments{
 For the color difference $\langle m-K_{20}\rangle$ we give the mean color and the dispersion in the
 color.  The statistical uncertainty in the mean color is generally less than 0.05~mag and the mean
 color was calculated over a magnitude range such that the survey magnitude limits would not affect
 the colors.
 We cannot calculate $\langle m-K_{20}\rangle$ for the ESP (too little overlap with the current 2MASS catalog),
 Century (no published object lists), and 2dFGRS (no published object lists) surveys.
 References: (1) Loveday et al. (1992); (2) Geller et al. 1997; (3) Marzke et al. 1994a; (4) Marzke et al. 1994b; 
 (5) Zucca et al. (1997); (6) Lin et al. 1996; (7) Bromley et al. 1998; (8) Marzke et al. 1998;  
 (9) Folkes et al. 1999.
   }
\end{deluxetable}

\begin{deluxetable}{rrrrr}
\label{tab-obslog}
\tablewidth{0pt}
\tablecolumns{5}
\tablecaption{Differential K$_s$-band Number Counts}
\scriptsize
\tablehead{
 \mc{K$_s$} &\mc{$\Delta$K$_s$}  &\mc{$N$}  &\mc{$\log(dN/dm)$}  &\mc{Poisson} \\
 \mc{(mag)} &\mc{(mag)}  &       &\mc{\#/mag/deg$^2$} &\mc{Errors} }
\startdata
        7.250    &0.50  &   20     &  $-$2.24\s   & 0.097 \\
        7.750    &0.50  &   36     &  $-$2.01\s   & 0.074 \\
        8.250    &0.50  &   53     &  $-$1.82\s   & 0.060 \\
        8.750    &0.50  &   84     &  $-$1.62\s   & 0.047 \\
        9.250    &0.50  &  172     &  $-$1.31\s   & 0.033 \\
        9.750    &0.50  &  320     &  $-$1.04\s   & 0.024 \\
       10.125    &0.25  &  298     &  $-$0.766    & 0.025 \\
       10.375    &0.25  &  439     &  $-$0.598    & 0.021 \\
       10.625    &0.25  &  635     &  $-$0.438    & 0.017 \\
       10.875    &0.25  &  872     &  $-$0.300    & 0.015 \\
       11.125    &0.25  & 1263     &  $-$0.141    & 0.012 \\
\enddata
\tablecomments{
   There are $N$ galaxies in each bin centered at K$_s$ and of width $\Delta$K$_s$,
   corresponding to number counts of $dN/dm$ in mag$^{-1}$~deg$^{-2}$ and its
   corresponding Poisson uncertainty. These are the 2MASS $\mu_{K_s}=20$~mag/arcsec$^2$
   circular isophotal magnitudes (Jarrett et al. 2000a).
   }
\end{deluxetable}

\end{document}